\documentclass[11pt]{article}
\usepackage{moriond,epsfig}

\def\be{\begin{equation}}
\def\ee{\end{equation}}
\def\bea{\begin{eqnarray}}
\def\eea{\end{eqnarray}}

\begin{document}
\vspace*{4cm}
\title{NEW COSMIC LOW ENERGY STATES OF NEUTRINO}

\author{E. I. GUENDELMAN AND A. B. KAGANOVICH }

\address{Physics Department, Ben Gurion University of the Negev,
\\
Beer Sheva
84105, Israel}

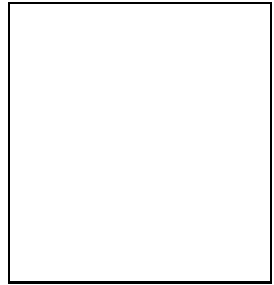
\maketitle\abstracts{ 
A field theory is studied where the consistency condition of equations of 
motion dictates strong correlation between states of "primordial"  fermion
fields and local value of the dark energy. 
In regime of the fermion densities typical for normal particle physics, the
primordial fermions split into three families identified with regular fermions.
When fermion energy density is comparable with dark energy density, the theory
allows transition to new type of states. The possibility of such Cosmo-Low Energy Physics
(CLEP) states is demonstrated in a model of FRW universe filled with 
homogeneous scalar field and
uniformly distributed nonrelativistic neutrinos. Neutrinos in CLEP state
are drawn into
cosmological expansion by means of dynamically changing their own parameters.
One of the features of the fermions in CLEP state is that in the late time
universe their masses
increase as $a^{3/2}$ ($a=a(t)$ is the scale factor). The energy
density of the cold dark matter consisting of neutrinos in CLEP state scales as
a sort of dark energy; this cold dark matter possesses negative pressure and for the
late time universe its equation of state approaches that of the cosmological
constant. The total energy density of such universe is less than it would be in
the universe free of fermionic matter at all.}

\section{Two Measures Theory (TMT)}
The TMT is a generally coordinate invariant theory
where the action has the form
   $ S = \int L_{1}\Phi d^{4}x +\int L_{2}\sqrt{-g}d^{4}x$
 including two Lagrangians $ L_{1}$ and $L_{2}$ and two
measures of integration: the usual one
$\sqrt{-g}$ and another $\Phi
=\varepsilon^{\mu\nu\alpha\beta}\varepsilon_{abcd}\partial_{\mu}\varphi_{a}
\partial_{\nu}\varphi_{b}\partial_{\alpha}\varphi_{c}
\partial_{\beta}\varphi_{d}$  built of four
scalar
fields
$\varphi_{a}$ ($a=1,2,3,4$); it is also a scalar density.
It is assumed that the Lagrangians $ L_{1}$ and $L_{2}$  are functions
of the matter
fields, the dilaton field $\phi$, the metric, the connection     
 but not of the
"measure fields" $\varphi_{a}$.
Solving equation that results from variation of $\varphi_{a}$, if 
$\Phi\neq 0$, we get
$ L_{1}=M^{4}$ where
$M$ is a constant of integration with the dimension of 
mass.
Important feature of TMT that is responsible for many
interesting and desirable results of the field theory models studied
so far~\cite{GK1}~-~\cite{GK6} 
 consists of the assumption that all fields, including
also metric, connection (or vierbein and spin-connection) and the
 measure fields $\varphi_{a}$ are independent dynamical variables.

In TMT there is no a need to postulate
the existence of three species for each type of fermions (like three neutrinos,
three charged leptons, etc.) but rather this is achieved as a dynamical effect
of TMT in normal particle physics conditions.
The matter content of our model includes the dilaton scalar
field $\phi$, two
so-called primordial fermion fields (the neutrino primordial field $\nu$
 and the electron primordial field $E$) and electromagnetic field $A_{\mu}$.
Generalization to the non-Abelian
gauge models including Higgs fields and quarks is straightforward~\cite{GK5}.
To simplify the presentation of the ideas we ignore also the chiral
properties of neutrino; this
can be done straightforward and does not affect the main results.

Keeping the general TMT structure of the action
 it is convenient to represent it in the following
form:
\begin{eqnarray}
S &=& \int d^{4}x e^{\alpha\phi /M_{p}}
(\Phi +b\sqrt{-g})\left[-\frac{1}{\kappa}R(\omega ,e)
+\frac{1}{2}g^{\mu\nu}\phi_{,\mu}\phi_{,\nu}\right]
\nonumber\\
&-&\int d^{4}x e^{2\alpha\phi /M_{p}}[\Phi V_{1} +\sqrt{-g}
V_{2}]-\int d^{4}x\sqrt{-g}  
\frac{1}{4}g^{\alpha\beta}g^{\mu\nu}F_{\alpha\mu}F_{\beta\nu}
\nonumber\\
&+&\int d^{4}x e^{\alpha\phi /M_{p}}(\Phi +k\sqrt{-g})\frac{i}{2}
\sum_{i}\overline{\Psi}_{i}  
\left(\gamma^{a}e_{a}^{\mu}\overrightarrow{\nabla}_{\mu}^{(i)}
-\overleftarrow{\nabla}_{\mu}^{(i)}
\gamma^{a}e_{a}^{\mu}\right)\Psi_{i}
\nonumber\\
&-&\int d^{4}x e^{\frac{3}{2}\alpha\phi /M_{p}}
\left[(\Phi +h_{\nu}\sqrt{-g})\mu_{\nu}\overline{\nu}\nu
+(\Phi +h_{E}\sqrt{-g})\mu_{E}\overline{E}E
\right]
\label{totaction}
\end{eqnarray}
where $\Psi_{i}$ ($i=\nu , E$) is the
general notation for the primordial fermion fields $\nu$ and $E$; 
 $V_{1}$ and $V_{2}$ are constants;
$F_{\alpha\beta}=\partial_{\alpha}A_{\beta}-
\partial_{\beta}A_{\alpha}$; \quad
   $\mu_{\nu}$ and
$\mu_{E}$ are  the mass parameters;
$\overrightarrow{\nabla}_{\mu}^{(\nu)}=\overrightarrow{\partial}_{\mu}+
\frac{1}{2}\omega_{\mu}^{cd}\sigma_{cd}$,
$\overrightarrow{\nabla}^{(E)}_{\mu}=\overrightarrow{\partial}_{\mu}+   
\frac{1}{2}\omega_{\mu}^{cd}\sigma_{cd}+ieA_{\mu}$;
$R(\omega ,e) =e^{a\mu}e^{b\nu}R_{\mu\nu ab}(\omega)$ is
the scalar curvature;
 $e_{a}^{\mu}$ and
$\omega_{\mu}^{ab}$ are the vierbein  and spin-connection;
$g^{\mu\nu}=e^{\mu}_{a}e^{\nu}_{b}\eta^{ab}$ and
$R_{\mu\nu ab}(\omega)=\partial_{\mu}\omega_{\nu ab}
+\omega_{\mu a}^{c}\omega_{\nu cb}
-(\mu\leftrightarrow\nu)$; constants $b, \, k, \, h_{i}$
are real dimensionless parameters.

The action (\ref{totaction}) is invariant
under
the global scale transformations
\begin{eqnarray}
    e_{\mu}^{a}\rightarrow e^{\theta /2}e_{\mu}^{a}, \quad   
\omega^{\mu}_{ab}\rightarrow \omega^{\mu}_{ab}, \quad
\varphi_{a}\rightarrow \lambda_{a}\varphi_{a}\quad
where \quad \Pi\lambda_{a}=e^{2\theta}
\nonumber
\\
A_{\alpha}\rightarrow A_{\alpha}, \quad
\phi\rightarrow \phi-\frac{M_{p}}{\alpha}\theta ,\quad
\Psi_{i}\rightarrow e^{-\theta /4}\Psi_{i}, \quad
\overline{\Psi}_{i}\rightarrow  e^{-\theta /4} \overline{\Psi}_{i}.
\label{stferm}
\end{eqnarray}

Except for a few special choices providing positivity of the
energy and the right chiral structure in the Einstein frame,
Eq.(\ref{totaction}) describes {\it
the most
general TMT action satisfying the formulated above symmetries}.

The appearance of a nonzero integration
constant $M^{4}$ in the mentioned above equation $ L_{1}=M^{4}$
 spontaneously breaks the scale invariance
(\ref{stferm}).
One can show that the measure $\Phi$ degrees of freedom
appear in all the equations of motion only through dependence
on the scalar
field $\zeta \equiv\Phi/\sqrt{-g}$.
In particular, the gravitational and all matter
fields equations of motion include noncanonical terms
proportional to $\partial_{\mu}\zeta$.
It turns out that with the set of the new variables 
$\tilde{e}_{a\mu}=e^{\frac{1}{2}\alpha\phi/M_{p}}(\zeta
+b)^{1/2}e_{a\mu}$, \quad
$\Psi^{\prime}_{i}=e^{-\frac{1}{4}\alpha\phi/M_{p}}
(\zeta +k)^{1/2}(\zeta +b)^{-3/4}\Psi_{i}$ \, ($\phi$ and
$A_{\mu}$ remain the same)
which we call the Einstein frame,
 the spin-connections become those of the
Einstein-Cartan space-time and the noncanonical terms proportional to
$\partial_{\mu}\zeta$ disappear from all equations of motion. 
   
The gravitational
equations
in the Einstein frame take the standard GR form
$G_{\mu\nu}(\tilde{g}_{\alpha\beta})=\frac{\kappa}{2}T_{\mu\nu}^{eff}$
where
$T_{\mu\nu}^{eff}=K_{\mu\nu}
+\tilde{g}_{\mu\nu}V_{eff}(\phi ;\zeta)+T_{\mu\nu}^{(em)}
+T_{\mu\nu}^{(f,can)}+T_{\mu\nu}^{(f,noncan)}$.
Here $K_{\mu\nu}=\phi_{,\mu}\phi_{,\nu}-\frac{1}{2}
\tilde{g}_{\mu\nu}\tilde{g}^{\alpha\beta}\phi_{,\alpha}\phi_{,\beta}$;
\quad
$G_{\mu\nu}(\tilde{g}_{\alpha\beta})$ is the
Einstein tensor
in the Riemannian space-time with the metric
$\tilde{g}_{\mu\nu}$; \quad
$V_{eff}(\phi ;\zeta)= (\zeta +b)^{-2}
\left[b\left(sM^{4}e^{-2\alpha\phi/M_{p}}+V_{1}\right)-V_{2}\right]$;\quad
$T_{\mu\nu}^{(em)}$ is the canonical energy momentum tensor for the
electromagnetic field;
$T_{\mu\nu}^{(f,can)}$ is the canonical energy momentum tensor for
(primordial) fermions $\nu^{\prime}$ and $E^{\prime}$ in
curved space-time
including also standard electromagnetic interaction of $E^{\prime}$.
$ T_{\mu\nu}^{(f,noncan)}=-\tilde{g}_{\mu\nu}\sum_{i}F_{i}(\zeta)
\overline{\Psi^{\prime}}_{i}\Psi^{\prime}_{i}
\equiv \tilde{g}_{\mu\nu}\Lambda_{dyn}^{(ferm)}$
\quad ($i=\nu^{\prime},E^{\prime}$)
is the {\em noncanonical} contribution
of the fermions into the energy momentum tensor,
and
$F_{i}(\zeta)\equiv
\mu_{i}2^{-1/2}(\zeta +k)^{-2}(\zeta +b)^{-1/2}
[\zeta^{2}+(3h_{i}-k)\zeta +2b(h_{i}-k)+kh_{i}]$.
The structure of
$T_{\mu\nu}^{(f,noncan)}$ shows that it is
originated by fermions but  behaves as
 a sort of  variable cosmological constant.
This is
why we will refer to it as {\it dynamical fermionic $\Lambda$
 term}
$\Lambda_{dyn}^{(ferm)}$. 
The dilaton $\phi$ field equation in the new variables
reads
$(-\tilde{g})^{-1/2}\partial_{\mu}
(\sqrt{-\tilde{g}}\tilde{g}^{\mu\nu}\partial_{\nu}\phi) -
2\alpha\zeta (\zeta +b)^{-2}M_{p}^{-1}M^{4}e^{-2\alpha\phi/M_{p}}=0$.
Equations for the primordial fermions
in terms of the   new
variables take the standard form of fermionic equations
in the Einstein-Cartan space-time  where the standard electromagnetic
interaction
 presents also. All the novelty consists of the form of
the $\zeta$ depending "masses" $m_{i}(\zeta)$ of the primordial fermions
$\nu^{\prime}$, $E^{\prime}$:
\begin{equation}
m_{i}(\zeta)= 
\frac{\mu_{i}(\zeta +h_{i})}{(\zeta +k)(\zeta +b)^{1/2}}
\qquad i=\nu^{\prime},E^{\prime}.
 \label{muferm1}
\end{equation}
It should be noticed that change of variables we have performed
provide also a conventional form
of the covariant conservation law of fermionic current
$j^{\mu}=\overline{\Psi}^{\prime}\gamma^{a}\tilde{e}_{a}^{\mu}\Psi^{\prime}$.

The scalar field $\zeta$ in the above equations is defined
by the constraint which is the consistency condition of
equations of motion.
In the Einstein frame it takes the form
\begin{equation}
-\frac{1}{(\zeta +b)^{2}}\left\{(\zeta
-b)\left[sM^{4}e^{-2\alpha\phi/M_{p}}+
V_{1}\right]+2V_{2}\right\}=
\sum_{i}F_{i}\overline{\Psi^{\prime}_{i}}\Psi^{\prime}_{i}.
\label{constraint3}
\end{equation}

Generically the constraint (\ref{constraint3}) determines $\zeta$ as
a very complicated
function of $\phi$, \,
$\overline{\nu^{\prime}}\nu^{\prime}$ and $\overline{E^{\prime}}E^{\prime}$ .   
However, there are  a few very
important particular situations where the theory allows exact
solutions of great interest~\cite{GK4}~,~\cite{GK5}.

\section{Fermion Vacuum, Regular Fermions, Fifth Force Problem}

 In the case of the complete absence of massive fermions
$\zeta = b-2V_{2}/
(V_{1}+sM^{4}e^{-2\alpha\phi/M_{p}})$ and the effective $\phi$-potential
is
\begin{equation}
V_{eff}^{(0)}(\phi)\equiv V_{eff}(\phi;\zeta)|_{\overline{\psi^{\prime}}\psi^{\prime}=0}
=\frac{(V_{1}+sM^{4}e^{-2\alpha\phi/M_{p}})^{2}}
{4[b(V_{1}+sM^{4}e^{-2\alpha\phi/M_{p}})-V_{2}]}.
\label{Veffvac}
\end{equation}
Assuming $bV_{1}>2V_{2}$ one can see that $V_{eff}^{(0)}$ monotonically
decreases (as $\phi\rightarrow\infty$) to the
positive cosmological constant
$\Lambda^{(0)}
=\frac{V_{1}^{2}}
{4(bV_{1}-V_{2})}$.

In a typical particle physics
situation, say detection of a single fermion,
the measurement implies a
localization of the fermion which is expressed in developing a very large
value of  $|\overline{\Psi^{\prime}}\Psi^{\prime}|$.
According to the constraint (\ref{constraint3}) this is
possible if   
$F_{i}(\zeta)\approx 0, \, i=\nu^{\prime},E^{\prime}$
(which gives two constant solutions for $\zeta$)
or
$\zeta \approx -b$.
These solutions allow to describe the effect of splitting
of the primordial fermions into three generations of the  
regular fermions (for details see ~\cite{GK4},~\cite{GK5}).
 It is interesting also
that for the first two generations (which we associate with the solutions
where $F_{i}(\zeta)\approx 0$) their coupling to the dilaton $\phi$
is automatically strongly suppressed
which provides a solution of the fifth force problem.

\section{Cosmo-Low Energy Physics (CLEP) states }

It turns out that besides the normal fermion vacuum
where the fermion contribution to the constraint is totally
negligible,
TMT predicts possibility of so far unknown
 states which can be realized,  for example,
in astrophysics and cosmology. Let us study a toy model~\cite{GK6}
 where
in addition to the
homogeneous scalar field $\phi$,  the spatially flat
universe is filled also with uniformly
distributed nonrelativistic neutrinos as a model of  dark matter.  
 Spreading of the neutrino
wave packets during their free motion lasting a long time yields
extremely small values of
 $\overline{\Psi}^{\prime}\Psi^{\prime}= u^{\dagger}u$
($u$ is the large component of the Dirac spinor $\Psi^{\prime}$).
There is a solution where the decaying fermion
contribution $u^{\dagger}u\sim \frac{const}{a^{3}}$ to the
constraint is compensated by
approaching $\zeta\rightarrow -k$. Then solving (\ref{constraint3})
for $\zeta$ we have to take into account both sides of the
constraint.
 After
averaging over typical cosmological scales (resulting in
the Hubble low), the constraint (\ref{constraint3})  reads
\begin{equation}
-(k+b)\left(sM^{4}e^{-2\alpha\phi/M_{p}}+V_{1}\right)
+2V_{2}+
(b-k)^{2}\frac{n^{(\nu)}_{0}}{a^{3}}
F_{\nu}(\zeta)|_{\zeta\approx -k}
=0
\label{constr-k}
\end{equation}   
where  $F_{\nu}(\zeta)|_{\zeta\approx -k}=  
\mu_{\nu}(h_{\nu}-k)(b-k)^{1/2}(\zeta +k)^{-2}+O((\zeta +k)^{-1})$
 and $n^{(\nu)}_{0}$ is a constant
determined by the total number of the cold neutrinos.

The crucial role in the CLEP solutions of the cosmological equations 
belongs to the neutrino
$\Lambda_{dyn}^{(ferm)}$ term.
We assume here that
$V_{1}>0, \enspace V_{2}>0 \quad and \quad
b>0, \enspace k<0, \enspace h_{\nu}<0, \enspace h_{\nu}-k<0,
 \enspace b+k<0 $. 
 In the late time universe,
the  pressure and density of the uniformly
distributed neutrinos in the CLEP state
 \begin{equation}
-P_{clep}=\rho_{clep}=\frac{2V_{2}+|b+k|V_{1}}{(b-k)^{2}}
+\frac{|b+k|}{(b-k)^{2}}
M^{4}e^{-2\alpha\phi/M_{p}}
\label{rho-dm-de}
\end{equation}
are typical for the dark energy sector including both
a cosmological constant and an exponential $\phi$-potential.
The total energy density and
the total pressure  
in the same regime are
$\rho^{(total)}_{dark}
 =\frac{1}{2}\dot{\phi}^{2}+U_{dark}^{(total)}(\phi)$, \quad
$P^{(total)}_{dark}
=\frac{1}{2}\dot{\phi}^{2}-U_{dark}^{(total)}(\phi)$,
where 
$U_{dark}^{(total)}(\phi)=\Lambda 
+\frac{|k|}{(b-k)^{2}}M^{4}e^{-2\alpha\phi/M_{p}}$ and
$\Lambda =\frac{V_{2}+|k|V_{1}}{(b-k)^{2}}$.
 The {\it remarkable result} is that
$V_{eff}^{(0)}(\phi)>U_{dark}^{(total)}(\phi)$
which means that (for the same values of $\dot{\phi}^{2}$)
{\it the universe in "the CLEP state" has a lower energy density
than in the fermion vacuum state}. One should
emphasize that this result does not imply at all that $\rho_{clep}$
is negative.

For
illustration of what kind of solutions one can expect, let us take the
{\em particular value} for the parameter $\alpha$, namely
$\alpha =\sqrt{3/8}$.
Then  the cosmological equations allow the following analytic solution
at the late time universe:
$\phi(t)=\frac{M_{p}}{2\alpha}\varphi_{0}+
\frac{M_{p}}{2\alpha}\ln(M_{p}t)$,\quad
$a(t)\propto t^{1/3}e^{\lambda t}$,
where
$\lambda =\frac{1}{M_{p}}\sqrt{\frac{\Lambda}{3}}$,
\quad $e^{-\varphi_{0}}=
\frac{2(b-k)^{2}M_{p}^{2}}{\sqrt{3}|k|M^{4}}\sqrt{\Lambda}$.
The mass of the neutrino in such CLEP state increases
exponentially in time 
$m_{\nu}|_{CLEP}\sim a^{3/2}(t)\sim
t^{1/2}e^{\frac{3}{2}\lambda t}$.

\end{document}